\documentclass[a4paper,11pt]{article}
\usepackage{aaskaiid}
\usepackage{orcidlink}

\usepackage{graphicx}	
\usepackage{amsmath}	
\usepackage{mathtools}

\usepackage{amssymb}	
\usepackage{caption}
\usepackage[table,xcdraw]{xcolor}
\usepackage{siunitx}
\usepackage{tabularx}   
\usepackage{array}      
\usepackage{booktabs}   

\usepackage{pdflscape}
      \newcolumntype{Y}{>{\raggedleft\arraybackslash}X} 
\usepackage{multirow} 
\usepackage{multicol}
\usepackage{multirow}
\usepackage{graphbox}
\usepackage{textcomp}
\usepackage{float}
\usepackage[caption = false]{subfig}
\usepackage{hyperref}
\usepackage{threeparttable} 

\usepackage{booktabs,tabularx,adjustbox}
\usepackage{siunitx}
\usepackage[table]{xcolor}
\usepackage{graphicx}   
\usepackage{pdflscape}  

\newcolumntype{C}{S[table-format=2.0]}
\newcolumntype{L}{>{\raggedright\arraybackslash}p{2.cm}}

\newcommand{\HWD}{2.em} 
\newcommand{\rotH}[1]{\makebox[\HWD][c]{\rotatebox[origin=l]{75}{\strut #1}}}

\title{Extending the SKA Across Africa:  The Case for a Continental African VLBI Network}

\ShortTitle{SKA Science with the African VLBI Network}

\author[1,2]{Emmanuel K. Bempong-Manful \orcidlink{0000-0002-1727-1224}}
\ShortName{Bempong-Manful et al.} 
\author[3,4]{Jompoj Wongphechauxsorn\orcidlink{0000-0002-7730-4956}}
\author[1]{Jack Radcliffe\orcidlink{0000-0002-0813-0497}}
\author[5]{Melvin Hoare}
\author[6]{Olga Bayandina \orcidlink{0000-0003-4116-4426}}
\author[7]{Pfesesani V. van Zyl \orcidlink{0000-0002-7510-6366}}
\author[8]{Cristina Garc\'ia-Mir\'o \orcidlink{0000-0003-1136-5016}}
\author[8]{Candela Chico \orcidlink{0009-0003-2595-0133}}
\author[9]{Naomi Asabre Frimpong\orcidlink{0000-0002-0623-003X}}
\author[4]{Hans-Rainer Kl\"ockner \orcidlink{0000-0002-0648-2704}}
\author[10]{Valente Cuambe}
\author[7]{Carla S. Mitchell}
\author[11]{Saul P. Phiri \orcidlink{0000-0003-0681-8738}}
\author[12]{Benedicta Woode \orcidlink{0000-0003-3078-3504}}
\author[12]{Emmanuel Proven-Adzri\orcidlink{0000-0002-9644-3134}}
\author[13]{Willice Obonyo\orcidlink{0000-0001-9038-1756}}
\author[14]{Tyler L. Bourke\orcidlink{0000-0001-7491-0048}}
\author[15]{Cristiana Spingola\orcidlink{0000-0002-2231-6861}}
\author[3]{Matthias Kadler \orcidlink{0000-0001-5606-6154}}
\author[16]{Ailing Wang \orcidlink{0000-0002-7351-5801}}
\author[17]{Roger P. Deane}
\author[15]{Isabella Prandoni \orcidlink{0000-0001-9680-7092}}
\author[15]{Rocco Lico \orcidlink{0000-0001-7361-2460}} 
\author[13] {James O. Chibueze\orcidlink{0000-0002-9875-7436}}
\author[1] {Robert Beswick}
\author[1] {Simon Garrington}
\author[18,19] {Benito Marcote}
\author[20,21] {Tao An \orcidlink{0000-0003-4341-0029}}
\author[15]{Marcello Giroletti}
\author[22,1]{Isaac M. Mutie\orcidlink{0000-0003-2024-3862}}
\author[23,4]{Anne-Kathrin Baczko\orcidlink{0000-0003-3090-3975}}
\author[24,2]{Johannes Allotey \orcidlink{0000-0001-6410-1001}}
\author[25]{Stefano Giarratana \orcidlink{0000-0002-2815-7291}}
\author[26]{Rasha M. Samir\orcidlink{0000-0003-2716-8332}}

\affiliation[1]{Jodrell Bank Centre for Astrophysics, Department of Physics and Astronomy, The University of Manchester, Manchester M13 9PL, UK}
\affiliation[2]{School of Physics, University of Bristol, Tyndall Avenue, Bristol BS8 1TL, UK}

\emailAdd{emmanuel.bempong-manful@manchester.ac.uk} 

\emailAdd{jompoj.wongphechauxsorn@uni-wuerzburg.de}

\affiliation[3]{Julius-Maximilians-Universit{\"a}t W{\"u}rzburg, Fakult{\"a}t für Physik und Astronomie, Institut für Theoretische Physik und Astrophysik, Lehrstuhl für Astronomie, Emil-Fischer-Str. 31, D-97074 W{\"u}rzburg, Germany}

\emailAdd{hrk@mpifr-bonn.mpg.de}
\affiliation[4]{Max-Planck-Institut f{\"u}r Radioastronomie, Auf dem H{\"u}gel 69, D-53121 Bonn, Germany}

\affiliation[5]{School of Physics and Astronomy, University of Leeds, Leeds LS2 9JT UK}

\affiliation[6]{SKA Observatory, 2 Fir Street, Black River Park, Observatory, 7925, Cape Town, South Africa}

\affiliation[7]{South African Radio Astronomy Observatory, Farm 502 JQ, Broederstroom road, Hartebeesthoek, 1740, South Africa}
\emailAdd{pvanzyl@sarao.ac.za}

\emailAdd{c.garciamiro@oan.es}
\affiliation[8]{National Astronomical Observatory, C. de Alfonso XII, 3, 28014 Madrid, Spain}

\emailAdd{naomi.asabrefrimpong@oao.iau.org}
\affiliation[9]{ IAU Office for Astronomy Outreach, National Astronomical Observatories Japan, Mitaka, Tokyo, Japan}

\emailAdd{valente.a.cuambe@a-raege-az.pt}
\affiliation[10]{RAEGE-Az, Caminho dos piquinhos s/n, Santa Marira, Portugal}

\affiliation[11]{Department of Physics, Copperbelt University, 21692 Jambo Drive, Kitwe, Zambia}
\emailAdd{saul.phiri6@cbu.ac.zm}

\affiliation[12]{ Ghana Space Science and Technology Institute, Ghana Atomic Energy Commission, P. O. Box LG 80, Legon, Accra, Ghana}
\emailAdd{Benedicta.woode@gaec.gov.gh}

\emailAdd{chibujo@unisa.ac.za}
\affiliation[13]{UNISA Centre for Astrophysics and Space Sciences (UCASS), College of Science, Engineering and Technology, University of South Africa, Cnr Christian de Wet Rd and Pioneer Avenue, Florida 1709, P.O. Box 392, 0003 UNISA, South Africa}
\emailAdd{tyler.bourke@skao.int}
\affiliation[14]{SKA Observatory, Jodrell Bank, Lower Withington, SK11 9FT, UK}

\affiliation[15]{INAF Istituto di Radioastronomia, via P. Gobetti 101, 40129, Bologna, Italy}

\affiliation[16]{Key Laboratory of Particle Astrophysics, Institute of High Energy Physics, Chinese Academy of Sciences, Beijing 100049, China}

\affiliation[17]{Wits Centre for Astrophysics, University of the Witwatersrand, 1 Jan Smuts Avenue, Johannesburg 2000, South Africa}

\emailAdd{marcote@jive.eu}
\affiliation[18]{Joint Institute for VLBI ERIC, Oude Hoogeveensedijk 4, 7991~PD Dwingeloo, The Netherlands}
\affiliation[19]{ASTRON, Netherlands Institute for Radio Astronomy, Oude Hoogeveensedijk 4, 7991~PD Dwingeloo, The Netherlands}

\emailAdd{antao@shao.ac.cn}
\affiliation[20]{Department of Astronomy, University of Science and Technology of China, Hefei, Anhui 230026, China}
\affiliation[21]{Shanghai Astronomical Observatory, Chinese Academy of Sciences, 80 Nandan Road, Shanghai 200030, China}

\affiliation[22]{Department of Astronomy and Space Science, Technical University of Kenya, PO Box 52428-00200, Nairobi, Kenya}

\emailAdd{anne-kathring.baczko@chalmers.se}
\affiliation[23]{Department of Physics and Astronomy, Chalmers University of Technology, SE-41296 Gothenburg, Sweden}

\affiliation[24]{STFC RAL Space, Rutherford Appleton Laboratory Harwell Campus, Didcot, Oxfordshire OX11 0QX, UK}

\affiliation[25]{INAF Osservatorio Astronomico di Brera, Via E. Bianchi 46, I-23807 Merate, Italy}

\emailAdd{rasha.samir@nriag.sci.eg}
\affiliation[26]{Department of Astronomy, National Research Institute of Astronomy and Geophysics (NRIAG), EL Marsad Street 1, Helwan, Cairo, Egypt}

\abstract{%

The African continent holds the key to unlocking the full potential of global Very Long Baseline Interferometry (VLBI). Strategic placement of radio telescopes across Africa provides the crucial north–south and intermediate baselines that are currently missing from the global VLBI network. This expansion will dramatically enhance imaging fidelity and resolution. In this chapter, we propose a vision for a \emph{continental} African VLBI Network (AVN) that will operate in close synergy with SKA-Mid, enabling transformational science across all cosmic scales. %
While only the Hartebeesthoek Radio Astronomy Observatory (HartRAO) in South Africa and the Ghana Radio Astronomy Observatory (GRAO) are currently operational, several partner countries are in the process of refurbishing or converting existing antennas. Here, we advocate for the expansion of this network through the deployment of a limited number of SKA-Mid–type telescopes across the continent, creating an “African arm” of SKA-VLBI. %
With a maximum baseline of $\sim$ 9000 km (from Rabat, Morocco to Cassis, Mauritius), the proposed continental facility will surpass for example, the resolution that will be achieved by the next generation Very Large Array (ngVLA) by $\sim$ 10 \% at similar observing frequencies and significantly enhance global VLBI coverage.
Beyond the scientific and technical gains, the AVN represents a unique opportunity for sustainable growth in human capital, education, and innovation across Africa. Developing and operating a continental VLBI array will train the next generation of engineers, data scientists, and astronomers, stimulate local industry, and inspire public engagement in science and technology. %
We outline the current status, challenges, and potential roadmap towards realizing this vision, and we highlight how a continental African VLBI network will position Africa at the forefront of global radio astronomy.
}

\begin{document}
\maketitle

\section{Introduction}
\label{sec:Intro}
Following early attempts by the University of Florida in 1964 to observe the decametric bursts from Jupiter using unconnected antennas with separate clocks  (cf. \citealt{Carr_1970}), %
and later successful fringe detection between two unconnected antennas separated by over 3000 km \citep{Broten_1967}, the technique of Very Long Baseline Interferometry (VLBI) has become one of the most powerful tools in observational astronomy. VLBI provides the sharpest view of the Universe, achieving angular resolutions of a few milliarcseconds at centimetre wavelengths (e.g. \citealt{Pushkarev_2012}; \citealt{Cheng_2025}) and even microarcsecond scales at submillimetre wavelengths (e.g. \citealt{EHTCollaboration2019}). The breadth of science enabled by VLBI is extraordinary — from detecting superluminal motion in relativistic jets, mapping accretion disks of supermassive black holes (SMBHs), tracing stellar explosions, and imaging atomic hydrogen outflows, to localising Fast Radio Bursts and producing the first images of black hole shadows (see  \citealt{Rezaei_2023} and references therein).
With VLBI expected to be a standard capability for both SKA-Low and SKA-Mid, the inclusion of African telescopes will be critical for achieving optimal baseline coverage. A continental array of SKA dishes across Africa would provide the essential north–south and intermediate baselines needed to enhance the sampling of the Fourier plane, greatly improving image fidelity for SKA-VLBI science.

Across continental Africa, approximately three dozen large satellite Earth-station antennas, initially constructed in the 1970s and 1980s, offer a valuable legacy infrastructure for conversion into operational radio telescopes. 
These legacy infrastructures inspired the concept of the SKA African Partner countries initiative (e.g. \citealt{Gaylard_2011}) — a partnership of eight sub-Saharan African countries (i.e., Botswana, Ghana, Kenya, Namibia, Madagascar, Mauritius, Mozambique and Zambia) initially identified as collaborators in the SKA-Mid project led by South Africa. The aim was to repurpose redundant satellite antennas in these eight SKA African Partner countries into radio telescopes (cf. Table \ref{tab:avn_pc}) at relatively low cost (e.g. \citealt{Gaylard_2011}), following successful examples in Australia (e.g. \citealt{McCulloch_2005}), Ireland (e.g. \citealt{Gabuzda_2005}), Japan (e.g. \citealt{Fujisawa_2002}), UK and elsewhere (see \citealt{Barbosa_2021} and references therein).

However, in practice, converting these antennas into functional VLBI stations has proven to be far more challenging than initially anticipated. Beyond the significant financial investment required, the process is constrained by a shortage of specialised technical expertise, limited local manufacturing and engineering capacity, and complex bureaucratic and institutional frameworks. For historical reasons, much of Africa has been at a disadvantage in developing large-scale scientific infrastructure, and national investment in astronomy has, until recently, been limited. Encouragingly, the advent of the SKA has begun to transform this landscape — sparking renewed interest from African governments, universities, and industries.

It is therefore timely to build upon this momentum and envision a broader, more coordinated approach. Here, we propose a new paradigm that extends beyond the original eight SKA African Partner countries: \emph{a continental network of SKA-class dishes across Africa designed for intermediate-baseline VLBI}. Such a facility would not only complement SKA-Mid and global VLBI science but also drive technological advancement, capacity building, and socio-economic development across the continent.

\begin{figure}[h]
    \centering
\includegraphics[width=1.0\textwidth]{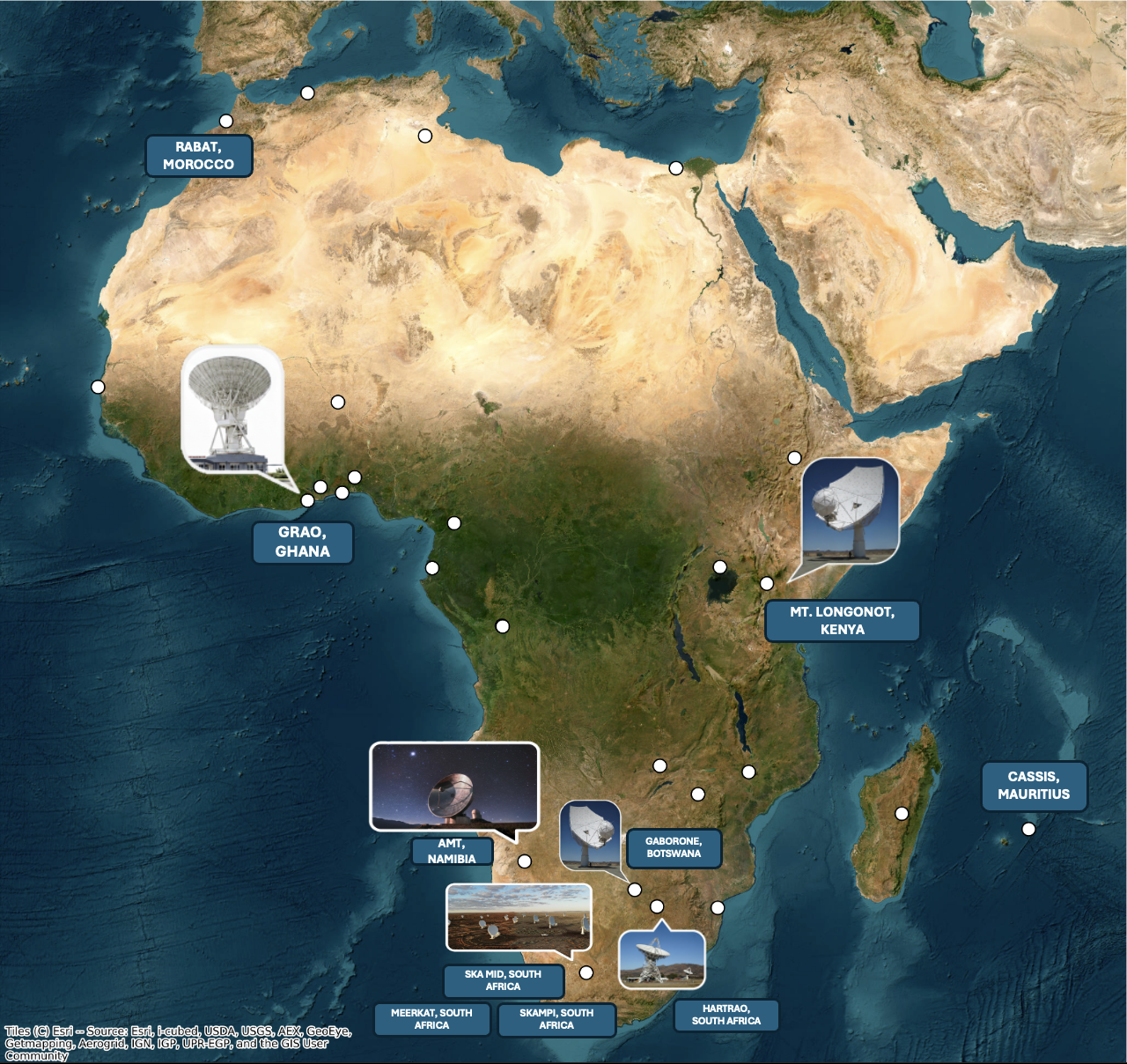}
    \caption{
    Overview of the locations of the decommissioned satellite antennas previously identified for inclusion in the SKA Partner countries array, together with the potential sites for new SKA-Mid–type antennas across Africa and other planned and current operational antennas. These locations illustrate a possible configuration of a future continental African VLBI array, with maximum baseline length between, e.g. Morocco and Mauritius. The full list of stations is provided in Table \ref{tab:avn_baseline}.}
    \label{fig:AVN-station-locations}
\end{figure} %

This chapter is organised as follows; in Section \ref{sec:ASU} we present an overview of the SKA Partner countries stations with a particular focus on the array status update. In Section \ref{sec:RnV} we present and describe our rationale and vision for a continental African SKA-Mid dishes for intermediate baseline VLBI and how this will impact the scientific research landscape. Sections \ref{sec:CCB} \& \ref{sec:SCPE} discuss the challenges that a continental African VLBI facility presents and the progress made in human capital development to date, and how science communication helps bridge these gaps. Finally, we summarise our proposition and provide some recommendations and plans for the future in Section \ref{sec:SCR}.

\section{The AVN Stations: Array Status Update}
\label{sec:ASU}
In this section, we provide a general review of the eight original AVN partner countries, with particular emphasis on the current and/or ongoing radio telescope infrastructure development, technical specifications, and operationality of the arrays -- following nearly two decades since the commencement of the AVN Partner countries initiative. For reference, we summarise the basic details about each station/antenna in Table \ref{tab:avn_pc}.

\subsection{South Africa: Hartebeesthoek Radio Astronomy Observatory (HartRAO)}

The Hartebeesthoek Radio Astronomy Observatory (HartRAO; \, e.g. \citealt{Nickola_2023}), operated by the South African Radio Astronomy Observatory (SARAO), is currently the only fully operational radio telescope in Africa participating in regular astronomical and geodetic VLBI sessions. It is a full member of both the European VLBI Network (EVN) and the International VLBI Service for Geodesy and Astrometry (IVS).

Located approximately 65 km northwest of Johannesburg, the facility hosts a 26-m Cassegrain antenna equipped with dual-polarized receivers covering the \emph{L} – \emph{K}-bands (1.7–24 GHz). The signal chain includes a hydrogen maser frequency standard, Digital Baseband Converter (DBBC2), and a Flexbuff recorder for high-speed buffering. A 10 Gb s$^{-1}$ data link supports near real-time e-VLBI and high-volume data transfer, ensuring full compatibility with international VLBI standards.

HartRAO continues to play a key role in both astronomical and geodetic VLBI and serves as a technical and training hub for the emerging African VLBI Network.

\subsection{South Africa: SKAMPI}

The SKA-MPIfR telescope (SKAMPI) is a 15-meter prototype antenna for the SKA-Mid located at the SKA site in the Karoo Desert, South Africa. It is funded  by the Max Planck Institute for Radio Astronomy (MPIfR)  and operated together with SARAO. 

Equipped with \emph{S}-band and \emph{Ku}-band receivers, SKAMPI was successfully tested in a VLBI observation involving the EVN and the Australian Long Baseline Array (LBA). Results from the fringe tests are presented elsewhere -- (cf. see chapter by \citealt{Wongphecauxson01.2026.SKA}). Due to its unique location, SKAMPI can be used as a substitute for SKA-VLBI if needed for the \emph{S}-band capability or low-sensitivity VLBI while maintaining SKA baselines.

\subsection{Ghana: Ghana Radio Astronomy Observatory (GRAO)}

The Ghana Radio Astronomy Observatory (GRAO; \, e.g. \citealt{Nsor_2024}), located at Kuntunse approximately 25 km northwest of Accra, is Africa’s second operational radio telescope. The 32 m antenna, originally constructed in 1979 by TIW Systems (Vertex–GDSatcom) under the INTELSAT program, served as a satellite Earth station until its decommissioning in 2009. It was identified in 2012 as a candidate for conversion under the SKA African Partner countries initiative to promote Africa’s participation in the SKA project.

The conversion process involved comprehensive refurbishment of the antenna’s mechanical, electronic, and optical systems. Obsolete control electronics and degraded servo components were replaced, while the structure underwent corrosion treatment, protective coating, and mechanical realignment. A redesigned azimuth cable wrap allows a total rotation of $\pm$\,305$^{\circ}$.

The beam-waveguide Cassegrain configuration employs four mirrors (two flat and two concave) aligned along a 12.68$^{\circ}$ waveguide axis. The dual-band receiver covers 5 GHz and 6.7 GHz, each with 128 MHz bandwidth and $\ge$\ 20 dB cross-polar isolation, supporting both continuum and spectral-line work.

A T4 Science iMaser3000 provides a frequency stability better than 5 $\times$ 10$^{-15}$ at 1000 s, meeting global VLBI requirements. Backend systems include a ROACH1-based spectrometer for single-dish operation and a DBBC2/Mark 5B$+$ system with 64 TB data storage for VLBI acquisition. Environmental monitoring is handled by a solar-powered MET4 system designed by HartRAO.

In 2017, a test observation with the EVN successfully detected reliable fringes, confirming stable baseband response and gain performance. This milestone validated the telescope’s technical readiness for international VLBI participation.

The successful transformation of the Kuntunse antenna marks a landmark achievement in African radio astronomy, demonstrating the feasibility of legacy conversions and providing a critical training platform for developing human capital in the region.

\begin{table*}
	\centering
	\begin{threeparttable}
		\begin{tabular}{l c c c c c c c c r} 
			\hline
			\hline
			{Country}
			&\multicolumn{1}{c}{Station Name}
			&\multicolumn{1}{c}{Antenna Diameter}
			&\multicolumn{1}{c}{Max. Operating Freq.}
			&\multicolumn{1}{c}{Status}\\

             & & [m] & [GHz] &  \\
			(1 ) & (2) & (3) & (4) & (5)  \\
			\hline
            Botswana........................ & BosRT & 15 & 6.8$^{\ast}$ & NS \\
			Ghana............................. & GRAO & 32 & 6.7 & CC \\    
			Kenya............................. & KRAO & 32 & 6.8$^{\ast}$ & NS \\
			Madagascar..................... & MadRT & 32 & 6.8$^{\ast}$ & NS \\
			Mauritius........................ & MauRT & 11 & 6.8$^{\ast}$ & NS \\
			Mozambique................... & MRAO & 7.6 & 6.8$^{\ast}$ & NS \\
			Namibia.......................... & AMT & 15 & 350$^{\ast}$ & NS \\
			South Africa (Lead)........ & HartRAO & 26 & 24 & OP \\
			Zambia............................ & KASERAO & 32 & 6.8$^{\ast}$ & NS \\
			
			\hline
		\end{tabular} \vspace{-0.04in}  
		\caption{%
		Station information for the original AVN Partner countries identified for the repurposing of redundant satellite antennas into radio telescopes. $^{\ast}$\,indicates proposed maximum operating frequency;\, CC: Conversion completed;\, NS: Not Started;\, OP: Operational. }%
        \label{tab:avn_pc}
\end{threeparttable}
\end{table*}

\subsection{Kenya: Kenya Radio Astronomy Observatory (KRAO)}

Efforts to establish a national radio astronomy observatory in Kenya initially focused on converting the decommissioned Longonot Earth Station’s 32\,m telecommunications antenna into a research-grade radio telescope, as part of the broader SKA African Partner countries initiative. However, subsequent technical assessments have shown that the Longonot facility has deteriorated significantly after years of disuse, with damaged infrastructure and obsolete control systems requiring extensive refurbishment. In addition, rapid urbanisation and the expansion of human settlements around Longonot have led to increased radio-frequency interference (RFI), further lowering the site’s suitability for scientific observations. 

Consequently, the focus of Kenya’s radio astronomy development has shifted toward capacity building and educational infrastructure. A key milestone in this transition is the recent installation of the Transient Array Radio Telescope (TART) at the Technical University of Kenya, a low-cost, wide-field instrument designed for training, skills development, and transient-source monitoring. The TART project, implemented with support from SARAO, the Development in Africa with Radio Astronomy (DARA) project, the Kenya Space Agency, and other international partners, has become central to developing local expertise in instrumentation, data processing, and observational techniques. Complementary site-testing campaigns are also underway to identify low-RFI regions suitable for a future observatory.

\subsection{Mozambique: Mozambique Radio Astronomy Observatory (MRAO)}

Mozambique occupies a strategic geographic position on the southeastern coast of Africa, bordered by the Indian Ocean. Its extensive coastline and stable geophysical environment provide favourable conditions for astronomical observations, particularly in radio astronomy. This potential, coupled with its location in the western Indian Ocean basin, makes Mozambique a promising site for the development of astronomical infrastructure.

Initial efforts to institutionalize astronomy in Mozambique commenced in the early 2000s and were significantly accelerated during the International Year of Astronomy in 2009. As noted by \citet{ribeiro_2009}, this period saw the implementation of foundational initiatives, including public outreach campaigns, curriculum development, and university-level training programs aimed at building local scientific capacity. These early activities were instrumental in positioning Mozambique as a potential host for a station of the SKA in Southern Africa, marking the country's initial engagement with major international astronomical projects.

Subsequent work by \citet{Barbosa_2013} detailed plans for the Mozambique Radio Astronomy Observatory (MRAO), a project designed to integrate research, education, and sustainable development. The MRAO constitutes a critical milestone in the nation's roadmap for radio astronomy and is intended to solidify its role within the AVN framework. 

Mozambique's participation in the AVN extends the east–west baseline coverage for interferometric observations in Africa and facilitates regional socio-economic development through technology transfer and specialized training (e.g. \citealt{alves_2019}). 

\subsection{Namibia: African Millimetre Telescope (AMT)}

The African Millimetre Telescope (AMT) is a planned 15 m radio telescope to be built on the Gamsberg Mountain in Namibia’s Khomas Highlands, about 30 km from the High Energy Stereoscopic System (H.E.S.S.) site. The project is a collaboration between Radboud University, the University of Amsterdam, the University of Oxford, the University of Namibia, the University of Turku, and the University of South Africa, supported by the European Research Council (ERC), the Netherlands Research School for Astronomy (NOVA), and partner universities.

The AMT’s primary goal is to join the Event Horizon Telescope (EHT) for millimetre-wavelength VLBI imaging of supermassive black holes and studies of black hole dynamics. The design covers ALMA Bands 1–6, with provision for a low-frequency receiver enabling centimetre-wave VLBI and single-dish observations. In single-dish mode, the telescope will be used for time-domain AGN monitoring, spectral-line studies of star-forming regions, and maser science (e.g. \citealt{AMT}).

By bridging millimetre and centimetre VLBI capability in Africa, the AMT will strengthen collaborations between global VLBI networks and SKA-Mid, and serve as a high-frequency complement to the AVN.

\subsection{Zambia: Mwembeshi and Kasempa Radio Astronomy Sites}

Zambia hosts a 32 m telecommunications antenna at the Mwembeshi Satellite Earth Station, located about 43 km west of Lusaka. Commissioned in 1974 and decommissioned in the early 2000s, the facility was later transferred to the SKA African Partner countries project through Zambia’s Ministry of Technology and Science. Although originally identified for conversion into a radio telescope, changes in project direction and technical constraints have delayed its integration into the SKA Partner countries innitiative. Nevertheless, the existing infrastructure may still serve as a foundation for a future data centre or support facility.

In addition, Zambia has designated a new 1 km$^{2}$ site for the Kasempa Radio Astronomy Observatory (KASERAO) in the Kasempa district of the North-Western Province, approximately 630 km from Lusaka. The site is expected to host future SKA antennas and related infrastructure.

Zambia remains an active SKA partner country, benefitting significantly from human capital development programmes through SARAO, DARA, PAPSSN, FAST4Future and PAP2SN. These initiatives have supported the training of Zambian scientists, engineers, and students in preparation for participation in both the African VLBI Network and SKA operations, marking important progress in building national capacity for radio astronomy.

\subsection{Botswana: Botswana Radio Astronomy}
The Botswana telescope will be SKA-Mid-type dish with similar system parameter as SKAMPI, although different receivers. The construction of the Botswana telescope (Boss) which will be located at the Botswana International University of Science and Technology will take place in 2026. This location will provide a short baseline (approximately 300 km) between HartRAO and Boss, while simultaneously offering a similar baseline between SKAMPI-HartRAO and SKAMPI-Boss. This enhanced setup will offer more tools for calibrations, thereby adding redundancy to the flux calibration process.

Madagascar and Mauritius are also known to have a history of radio astronomy. The design and siting of potential new telescopes in these two countries is driven by a combination of the broader science case of the SKA, the need for compatibility with existing networks and converted antennas, the provision of services, minimisation of radio frequency interference, and of course, costs (e.g. \citealt{Gaylard2013}).

\begin{figure}[h!]
    \centering
\includegraphics[width=1.0\columnwidth]{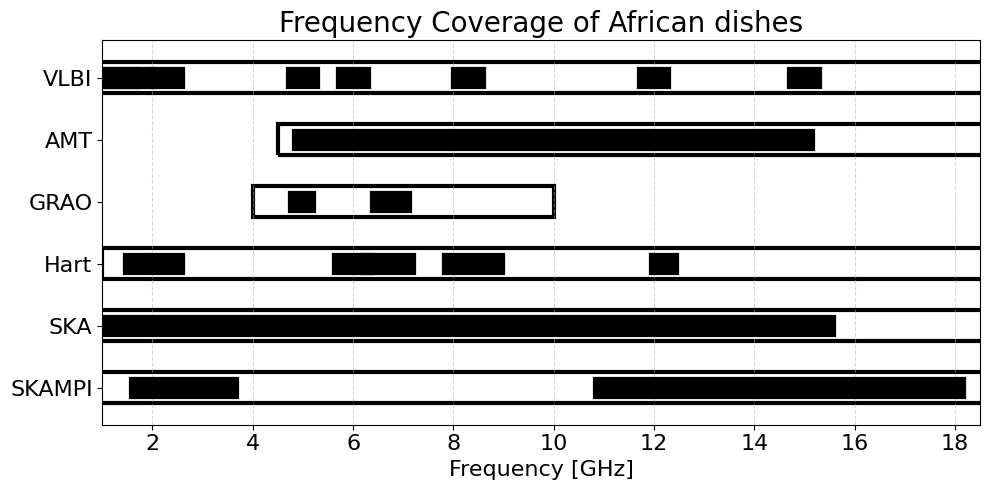} \\
    \caption{Frequency coverage of current African radio telescopes and SKA-Mid below 18 GHz compared with standard VLBI observing frequencies provided by EVNplanobs. The VLBI bands are shown as 256-MHz–wide boxes for reference. The boxes indicate possible frequency ranges for each telescope. The plot highlights existing overlaps as well as the expanded coverage achievable with future SKA-Mid–compatible receivers deployed across Africa.
    }
\label{fig:avn_freq_coverage}
\end{figure}

\section{Towards a Continental African SKA-Mid VLBI Network}
\label{sec:RnV}

\subsection{Rationale and Vision}
The scientific success of the SKA depends not only on the sensitivity of its core arrays but also on the ability to extend its resolving power through Very Long Baseline Interferometry (VLBI). Africa’s geography offers a unique and unparalleled opportunity to enhance global VLBI by providing critical north–south and intermediate baselines. To fully exploit this potential, we propose the deployment of a small number of SKA-Mid–type antennas in selected African countries (cf. Table \ref{tab:avn_baseline}) — with baselines extending from Rabat, Morocco to Cassis, Mauritius.

These stations would be built using the same 15-m dish design as the SKA-Mid antennas in South Africa, equipped with receivers covering identical frequency bands to ensure full compatibility for co-observing operations. This approach offers clear advantages: streamlined engineering and maintenance, seamless data integration with the SKA-Mid correlator, and shared hardware and software control systems.

Together, these new stations would form a truly continental VLBI network extending from North Africa to the Indian Ocean and spanning up to $\sim$\,9000 km (cf. Figure \ref{fig:AVN-station-locations}). Strategically positioned along both meridional and equatorial axes, the new VLBI network would complement SKA-Mid by providing intermediate baselines (hundreds to thousands of kilometres), bridging the gap between intra-array and intercontinental VLBI spacings.

\subsection{Scientific Impact}

\begin{figure*}
\hspace{2.5cm}\small{12GHz simulation of current and future stations} 
\\ 
    \centering 
      \includegraphics[width=0.498\textwidth]{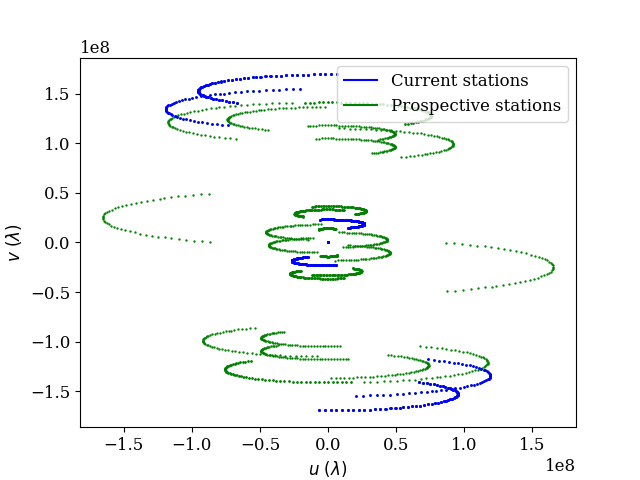} \hspace{-0.05in} 
    \includegraphics[width=0.498\textwidth]{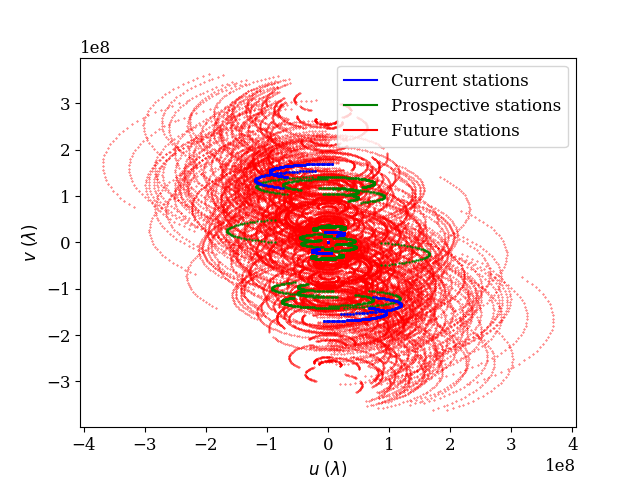}\\
    \vspace{0.15in}
    \includegraphics[width=0.330\textwidth]{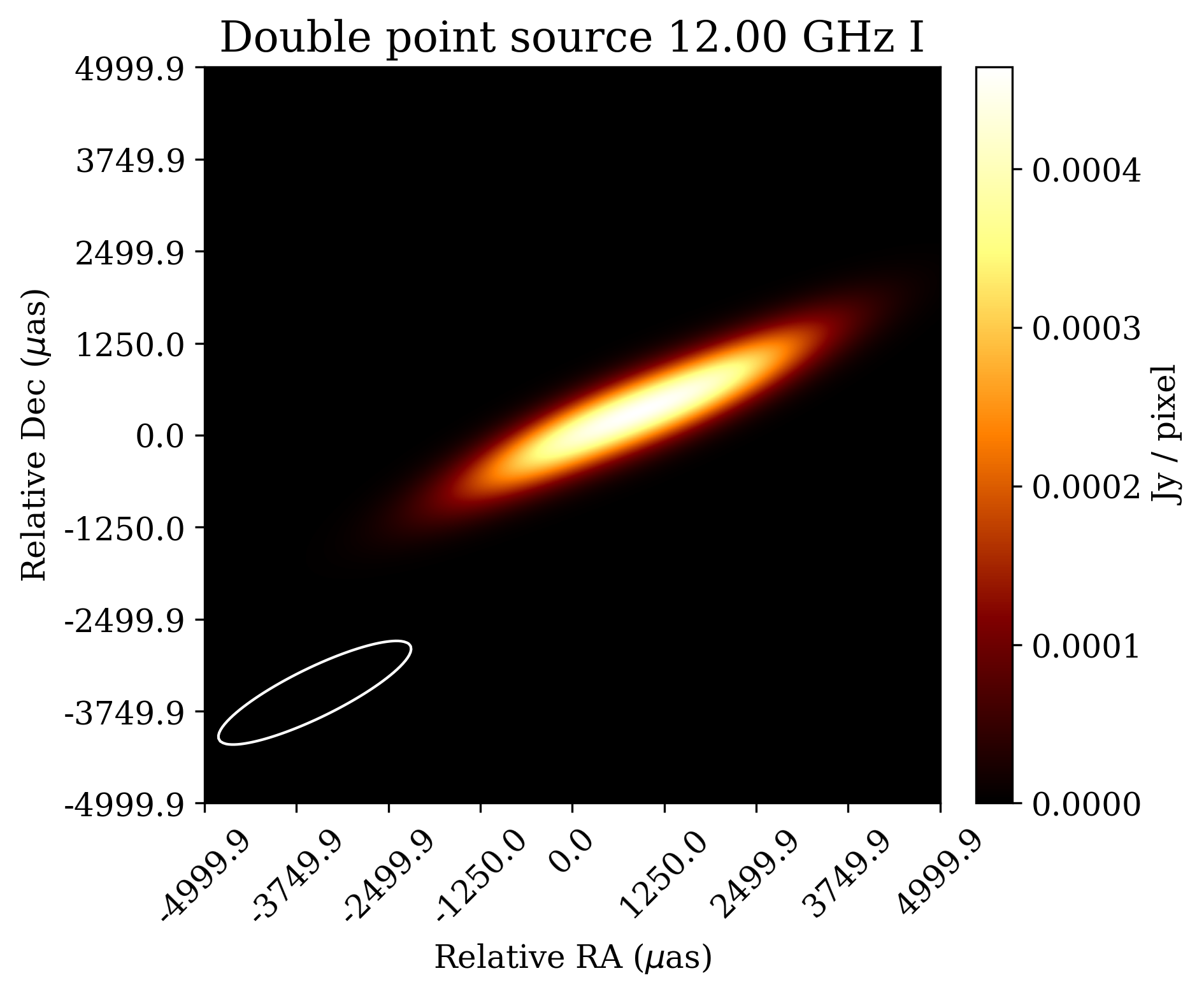} \hspace{-0.05in} 
    \includegraphics[width=0.330\textwidth]{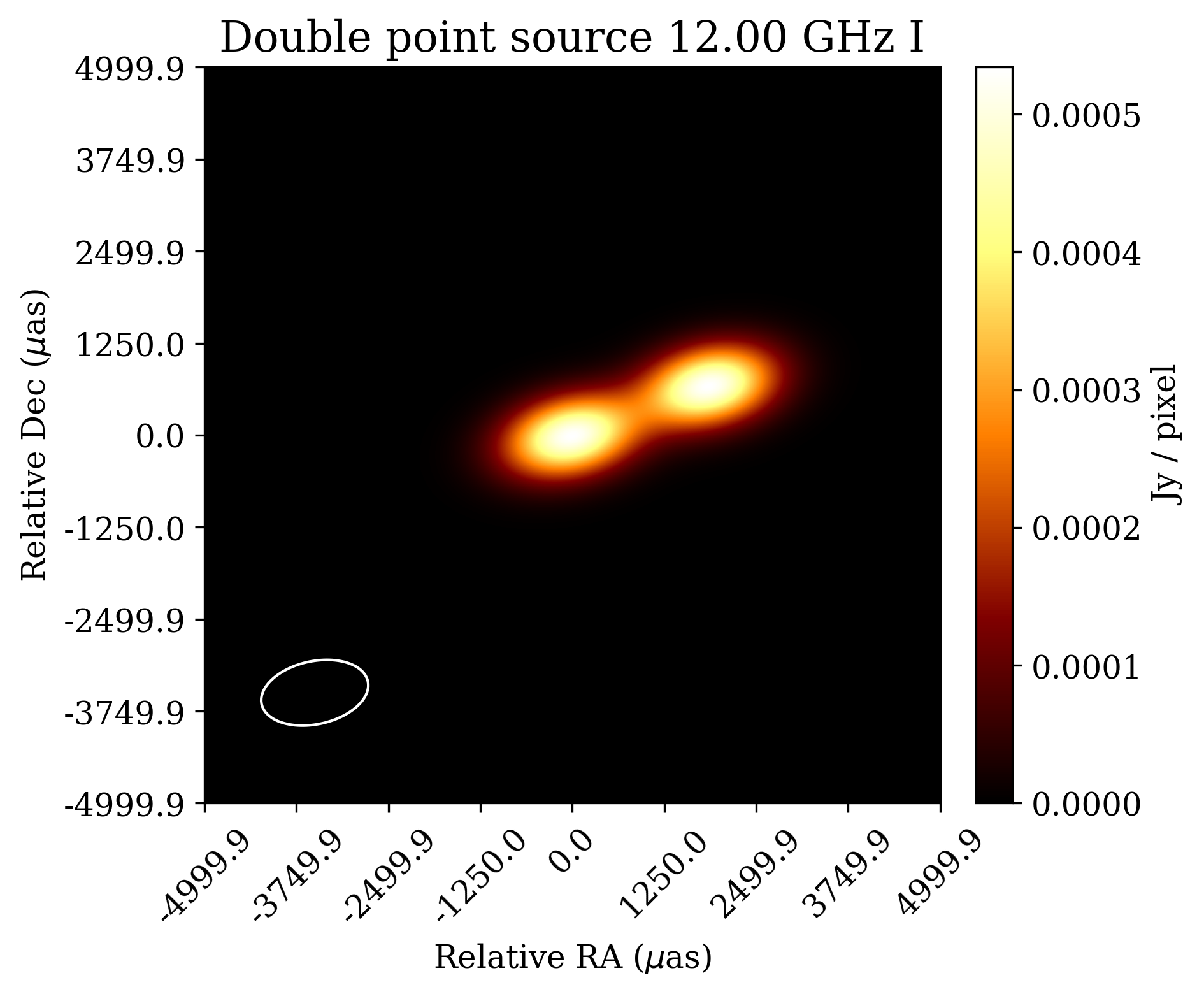}
    \includegraphics[width=0.330\textwidth]{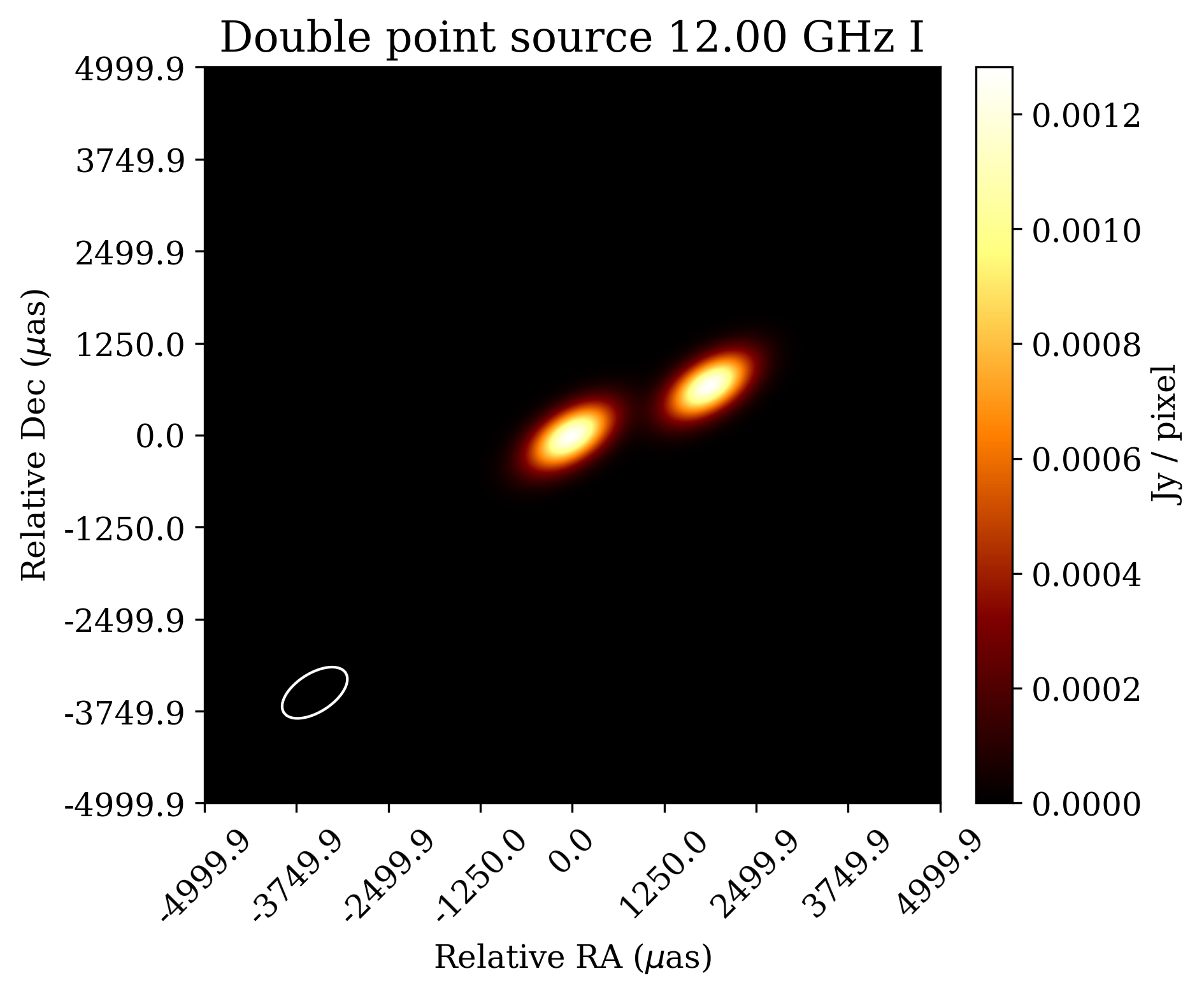}
    \caption{Top (left): \emph{uv}-coverage comparison between SKA-Mid and AVN-remote stations. Top (right): \emph{uv}-coverage comparison between current stations (blue), prospective stations (green), and future stations (red). Bottom: Results from simulation of a double point source at 12\,GHz with 1.5 mas angular separation; Left panel -- simulation results from current stations; Middle panel -- simulation results from prospective stations; Right panel -- simulation results from future stations. The results show that the addition of stations (i.e. Kenya, Botswana, and AMT), significantly improves the resolution and help separate the components. Current stations consist of SKA-Mid, SKAMPI, HartRAO and GRAO. Prospective stations are current stations with the Botswana telescope, the Kenya telescope, and AMT, while future stations are listed in Table \ref{tab:avn_baseline}.}
    \label{fig:uv-coverage}
\end{figure*}

From differential VLBI astrometry to large field of view (FoV) VLBI surveys, the breadth of science that will be enabled by SKA-VLBI has been extensively described elsewhere — see \citet{Paragi_2015} and references therein. Adding the proposed continental African VLBI array to the SKA is expected to increase by a factor of 10 the sensitivity and resolution of a typical 24-hour observation, using, for example, 26 African stations combined with a phased-up SKA-Mid array (cf. Figure \ref{fig:uv-coverage}).
The inclusion of these SKA-Mid–type antennas across Africa would have a transformative impact on VLBI science in several ways:

\begin{enumerate}
    {\item \bf Enhanced North–South Baselines:}\\
    Most current global VLBI arrays are dominated by east–west baselines, limiting imaging fidelity for equatorial and southern sources. Africa’s unique geography provides long north–south baselines — from Morocco and Egypt in the north to Madagascar and Mauritius in the south — improving Fourier-plane coverage and enabling more accurate imaging of complex astrophysical structures.
    
    {\item \bf Improved Imaging of Southern Sources:}\\
    The new African stations would allow high-resolution VLBI imaging of southern-sky targets — including the Galactic Centre, Magellanic Clouds, and nearby AGN — that are only partially accessible from existing northern-hemisphere networks. Together with SKA-Mid, these antennas would significantly enhance sensitivity and angular resolution for southern hemisphere science.
    Moreover, SKA-Mid$-$VLBI wide-field observations could extend census of many more calibrator-quality objects in the Southern sky (e.g. \citealt{Petrov2019}).

    {\item \bf Intermediate Baselines for SKA-VLBI:}\\
    The proposed array would provide baselines between a few hundred and a few thousand kilometres — a crucial range for connecting SKA-Mid to the global VLBI network. These baselines improve \emph{uv}-coverage, reduce sidelobe confusion, and bridge the spatial resolution gap between SKA-Mid interferometry and long-baseline VLBI.

    {\item \bf Long Baselines and Resolution Gains:}\\
    Extending from Morocco to Mauritius, the continental array would achieve baselines of up to 9000 km, delivering angular resolutions comparable to or exceeding those of currently operational VLBI arrays. This would make the African network a powerful global VLBI partner for studying compact sources, relativistic jets, and transient phenomena, among others (cf. see science chapters by \citealt{Baczko01.2026.SKA}, \citealt{Giarratana01.2026.SKA}, \citealt{Kadler01.2026.SKA}, \citealt{Panessa01.2026.SKA}).

    {\item \bf Extended Frequency Coverage:}\\
    By using SKA-Mid–standard receivers, the proposed African antennas would add critical support for Band 1 (0.35–1.05 GHz), overlapping with VLBI UHF band,  — a frequency range currently underrepresented in global VLBI. Only a few telescopes worldwide operate in this band, yet it provides the essential bridge between low-frequency VLBI with SKA-Low and the higher-frequency VLBI traditionally used in existing arrays. Expanding Band 1 participation across Africa would significantly broaden VLBI’s usable frequency range (cf. Figure \ref{fig:avn_freq_coverage}), enabling new studies of spectral turnovers in AGN and absorption features in intervening media.

    {\item \bf New Opportunities for Time-Domain and Multi-Messenger Astronomy:}\\
    Enhanced sensitivity and resolution would support monitoring of variable sources such as AGN, X-ray binaries, and masers, while precise localisation of FRBs and gravitational-wave counterparts would benefit directly from the array’s intermediate baselines and low-frequency capability.
\end{enumerate}

Beyond their contribution to global VLBI, the proposed African VLBI SKA-Mid dishes will serve as versatile single-dish observatories during non-VLBI observing periods. These telescopes can participate in monitoring programmes of variable radio sources, spectral-line and continuum surveys, and geodetic sessions, providing valuable standalone scientific data. Their high sensitivity and modern instrumentation will make them ideal for time-domain studies and rapid follow-up observations of transient events detected by SKA-Low or SKA-Mid. Moreover, their distributed locations across the continent will enable coordinated multi-frequency campaigns and long-term monitoring that are difficult to achieve with existing facilities. Equally important, these antennas will serve as training and education platforms, giving students and early-career researchers across Africa direct access to cutting-edge instrumentation and hands-on experience in radio astronomy operations, data analysis, and instrumentation — laying the foundation for the next generation of African astronomers and engineers.

\subsection{Technical and Infrastructural Considerations}

The use of SKA-Mid standard dishes ensures a high level of engineering uniformity and operational reliability. The antennas would employ the same cryogenic receiver systems and backend architecture, enabling data acquisition and correlation with minimal additional development. Co-observation with SKA-Mid and existing African facilities (e.g. HartRAO and GRAO) would be supported through established e-VLBI protocols and high-capacity fibre links.

Site selection will prioritise locations with existing telecommunications infrastructure, good sky coverage, and stable environmental conditions. Each proposed host country already engages in SKA-related activities, providing a foundation for technical support, data transport, and local expertise development.

By implementing SKA-Mid–compatible infrastructure across the continent, this initiative would create an integrated platform for scientific and technical collaboration within the SKA-VLBI framework. The resulting array would enable the next stage of high-resolution radio astronomy from Africa, forming the foundation for future developments in instrumentation, operations, and training discussed in subsequent sections.

\section{Challenges and Capacity Building Towards a Connected African VLBI Network}
\label{sec:CCB}

\subsection{Development of a Technical and Human Capital}

The development of a continental African VLBI network presents both exciting opportunities and significant challenges. Although several African partner countries host legacy satellite antennas suitable for conversion, only a few — such as South Africa and Ghana — have completed or nearly completed refurbishment. In many others, the process has proven complex and costly, requiring sustained investment, specialized expertise, and coordinated governmental support. The installation of new SKA-Mid–type antennas will face similar hurdles: from establishing technical readiness and maintenance capacity to ensuring the availability of trained engineers, scientists, and observatory staff.

Equally important are the challenges of data transport, correlation and archiving. While modern VLBI systems now rely on flexible software correlators, their operation demands substantial computing power, high-bandwidth fibre links, and robust data management frameworks. 
To build a robust digital infrastructure capable of handling the exascale data volumes expected from the AVN, it is essential to foster regional coordination and establish strategic partnerships with global data centres and existing international VLBI facilities. This will ensure seamless data integration and sustainable infrastructure across multiple African nations. 
The technical integration of new antennas is therefore only the first step — transforming them into a coherent, scientifically productive network will depend on long-term investment in both people and systems.

Despite these challenges, Africa’s engagement with the SKA has already demonstrated the transformative power of human capital development. Initiatives such as DARA \citep{Hoare2018} and DARA Big Data have trained hundreds of students and professionals in radio astronomy, engineering, and data science, linking these skills to broader sectors such as remote sensing, telecommunications, and medical imaging.
More recently, astronomy schools specializing in VLBI such as the first \href{https://www.sarao.ac.za/events/sarao-african-vlbi-school-2025/}{SARAO African VLBI School} and the \href{https://jive.eu/jvs2025}{JIVE VLBI School} brought in Pretoria students, post-docs and academic staff from South Africa and the African SKA Partner countries.
These programmes have fostered a generation of young African scientists and technologists who are now poised to drive the next phase of continental collaboration.

The growing network of space and data hubs envisioned around a continental Africa and SKA infrastructure offers an opportunity to align scientific progress with economic innovation. Co-located facilities — combining radio telescopes, satellite ground stations, and data centres — can anchor regional development while ensuring that Africa remains an active and self-sustaining partner in global radio astronomy. With sustained commitment, the creation of an integrated continental African VLBI facility will not only extend the reach of SKA-Mid and global VLBI but also serve as a model for how frontier science can catalyse technological and societal growth across an entire continent.

\subsection{Requirement for an African Correlator Hub}

With the growing number of radio telescopes being established across the African continent as part of the continental VLBI facility, there is an increasing need for a dedicated infrastructure to coordinate and support scientific and technical operations. The establishment of a continental VLBI correlation hub would serve as a central facility for processing, analyzing, storing, and managing data collected from the various telescopes within the network. Once four or more telescopes are operational, synthesis imaging of more complex astronomical structures becomes feasible, and at this stage, centralized operations and data management are essential to ensure consistency, accuracy, and efficiency in data processing.
The correlation facility would be responsible for real-time, offline, and archival correlation of data, using dedicated software correlators such as DiFX or specialized hardware correlators where appropriate. It would also host custom-built correlation pipelines and applications tailored to the specific requirements of the continental VLBI stations. This facility would integrate advanced beamforming and calibration tools to support both scientific observations and technical experiments aimed at improving the coherence and quality of combined data streams. Through these processes, the centre would provide calibration solutions to correct for instrumental and environmental effects, ensuring that the final correlated data meet international standards of precision and reliability.
In addition to its technical functions, the hub would provide a secure and scalable data storage infrastructure, enabling efficient handling of raw, correlated, and processed datasets. It would manage long-term archiving and facilitate data access and sharing among participating institutions and international collaborators. By maintaining a comprehensive database of metadata and observational records, the centre would ensure that all scientific data produced within the continental African VLBI network remains discoverable and reusable for future research.
Another critical role of the facility would be to provide extensive user support and training. It would offer technical assistance, user documentation, and hands-on guidance to astronomers, engineers, and data scientists involved in African VLBI operations. Regular training workshops and capacity-building programs would help develop expertise in VLBI data processing, calibration, and interpretation across the continent. Furthermore, the centre would function as a technology testbed for the development and validation of new correlation techniques, beamforming applications, and calibration algorithms.
Overall, the continental African VLBI correlator hub would not only serve as the operational and scientific backbone of the African VLBI Network but also as a strategic platform for advancing radio astronomy research and technical capacity in Africa. By enabling centralized data handling and facilitating collaboration with global VLBI networks, it would significantly enhance Africa’s contribution to international radio astronomy and provide a foundation for innovation in data science and engineering within the region.

\begin{figure*}
\hspace{2.5cm}
\small{6.5GHz simulation} \\
    \centering 
    
    \includegraphics[width=0.3\textwidth, trim=0cm 0cm 3cm 0cm, clip]{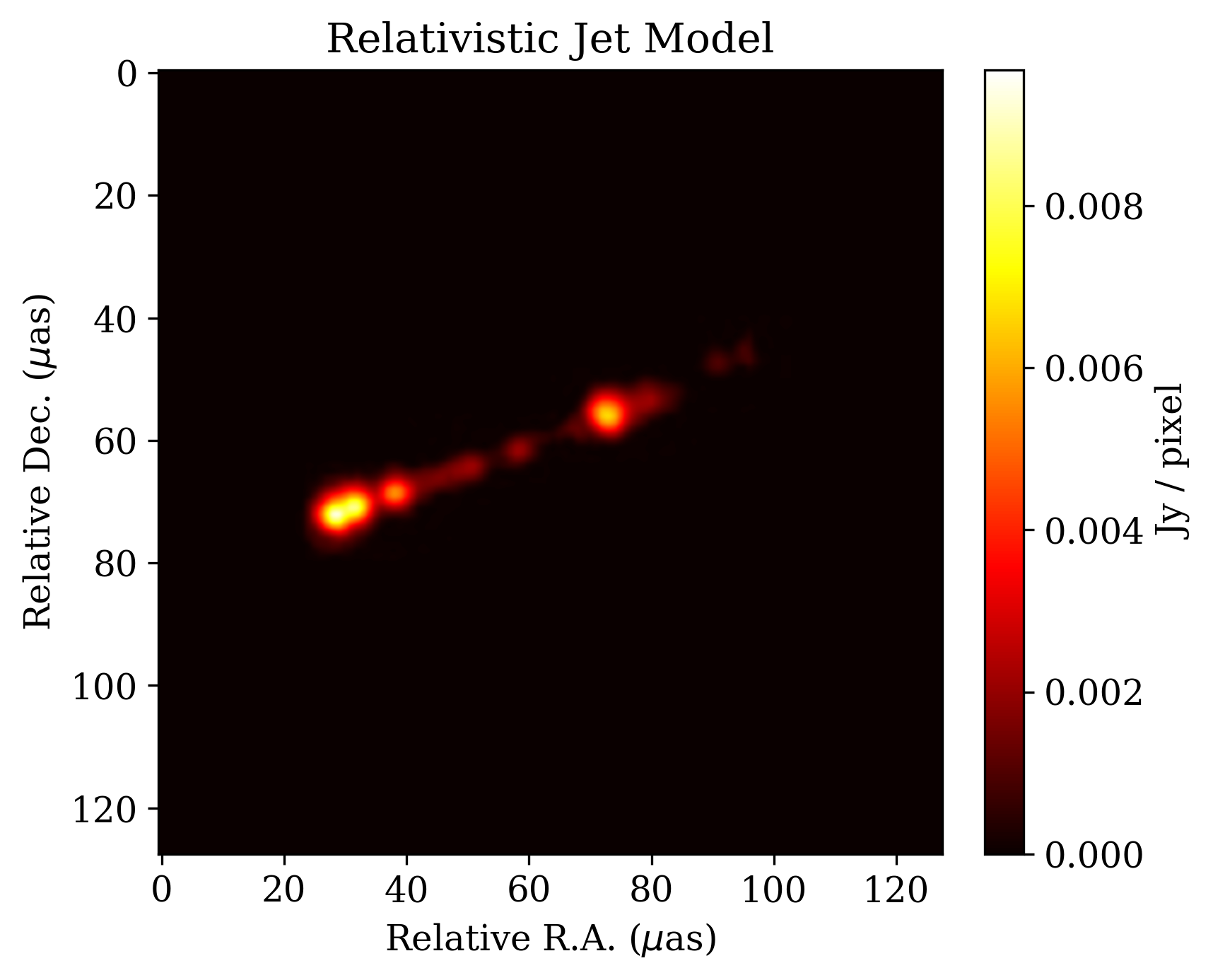}\\
    \includegraphics[width=0.25\textwidth]{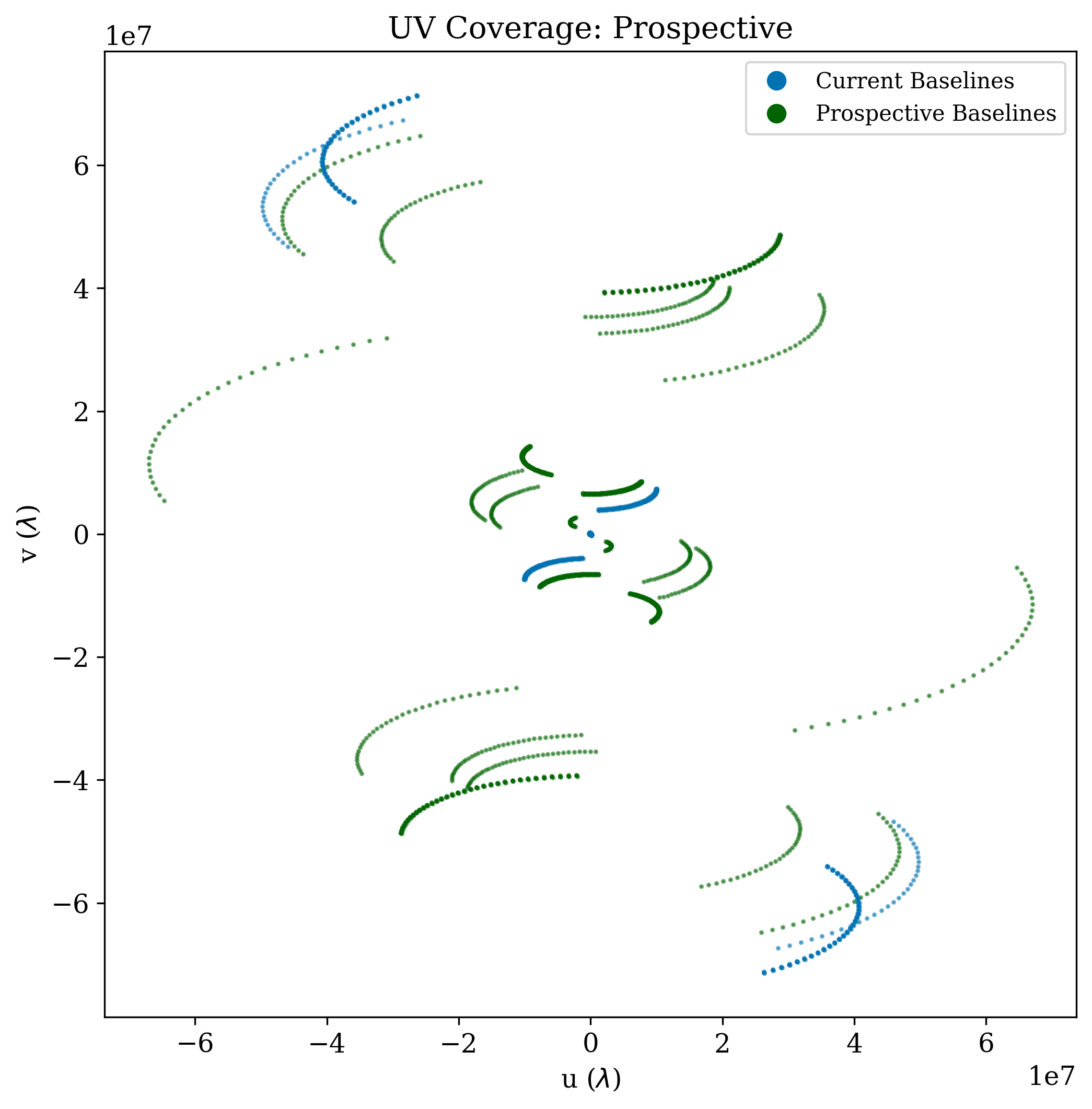} 
    \includegraphics[width=0.26\textwidth]{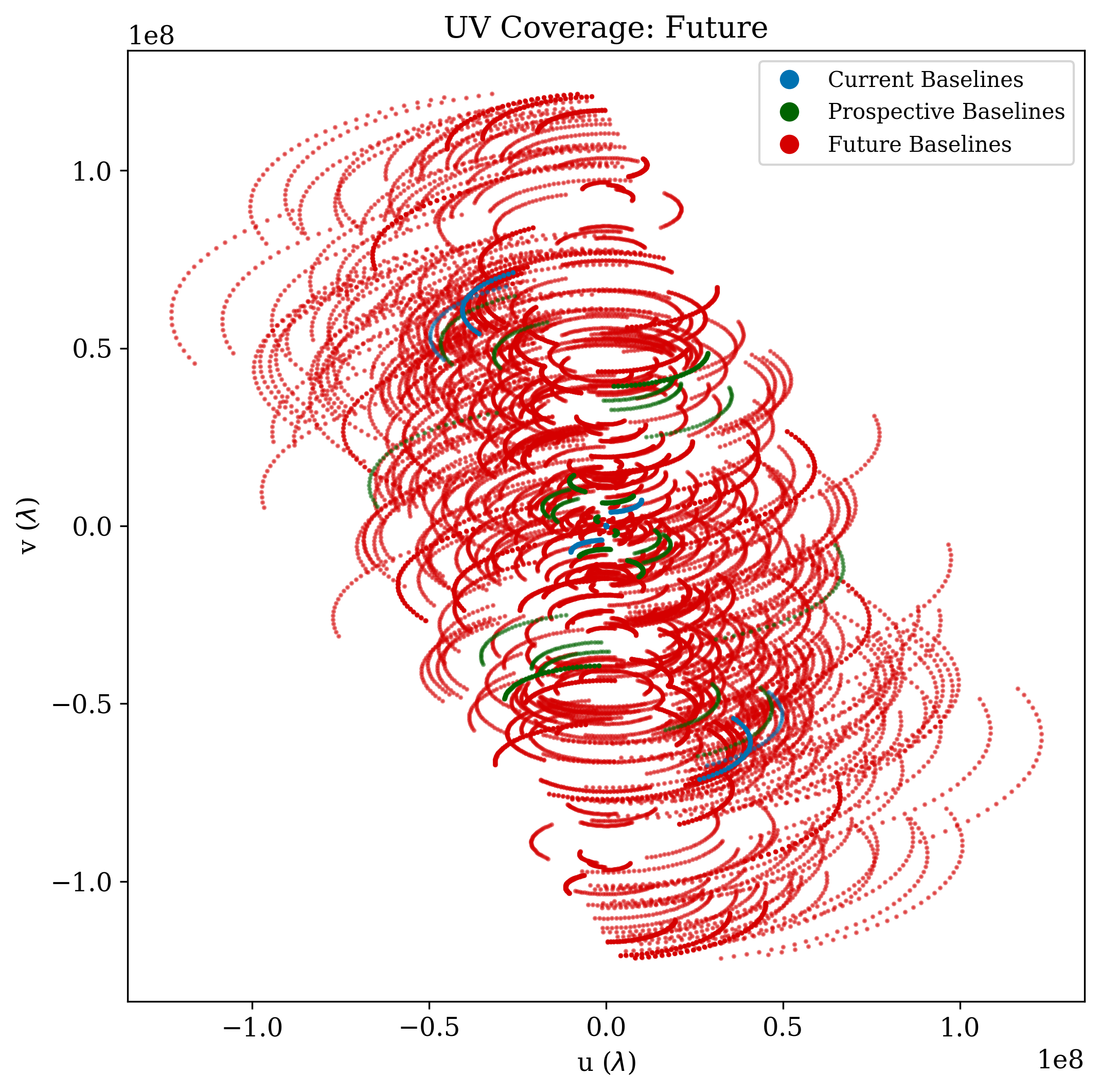}\\
    \includegraphics[width=0.33\textwidth]{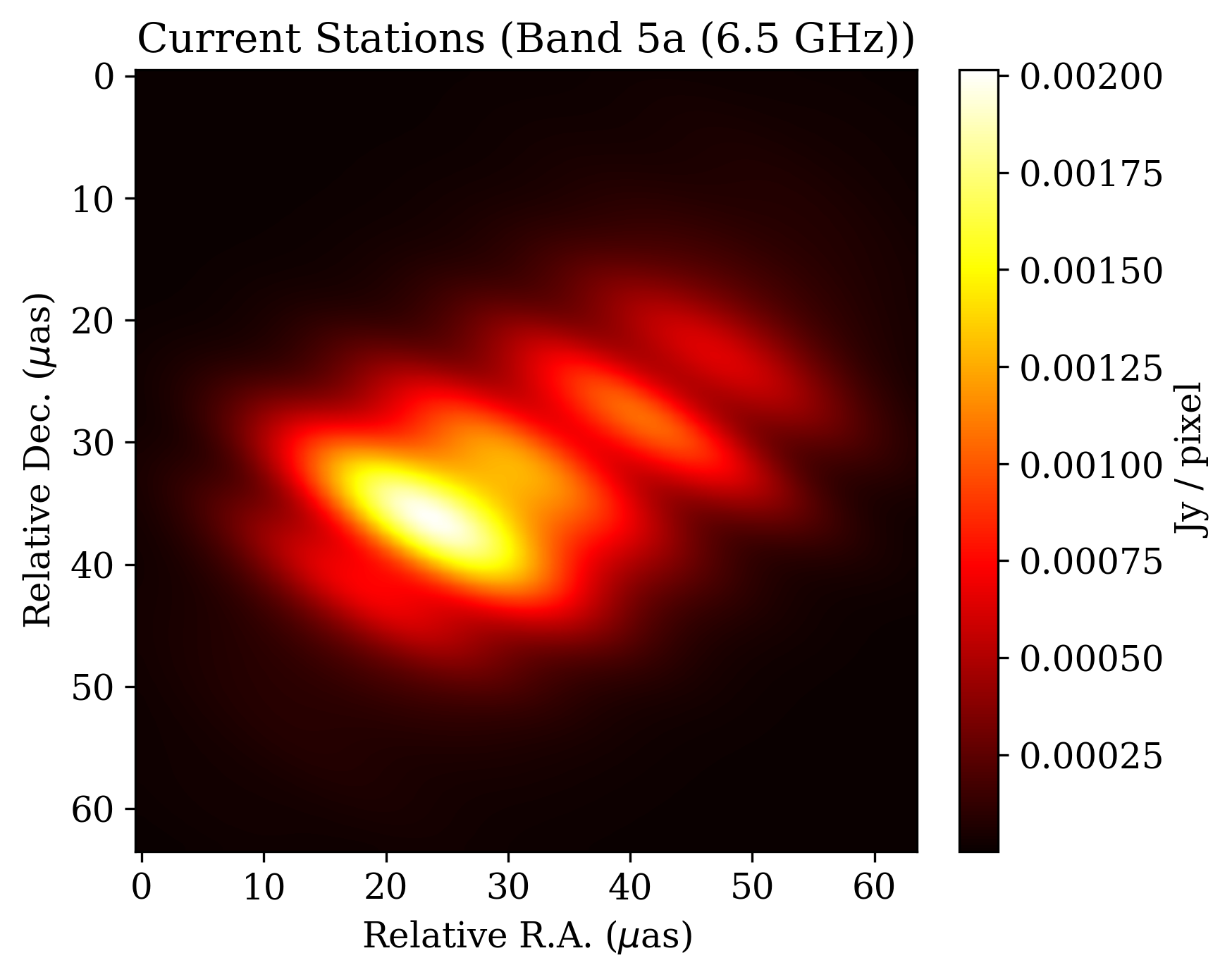} \hspace{-0.05in} 
    \includegraphics[width=0.33\textwidth]{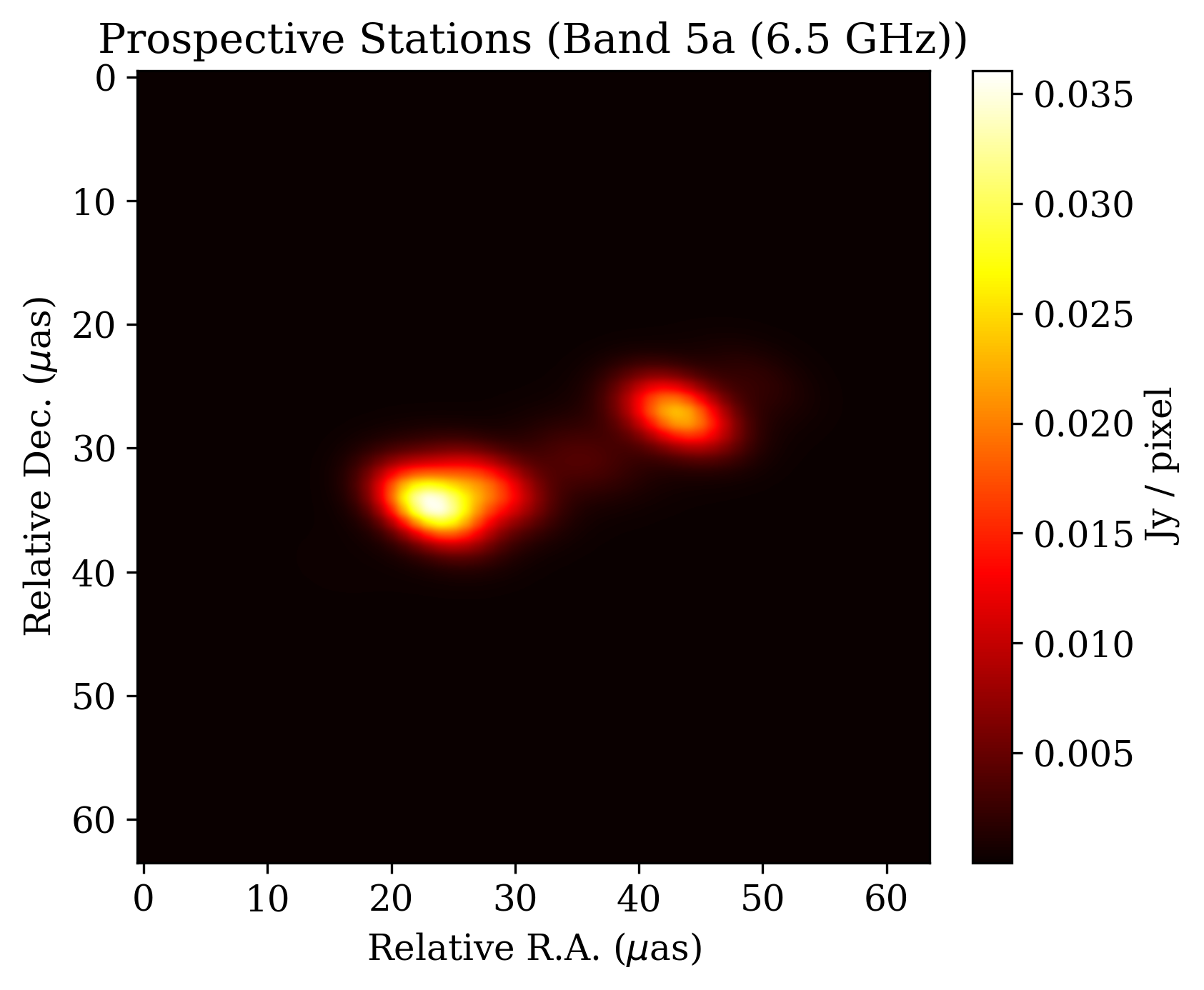}
    \includegraphics[width=0.33\textwidth]{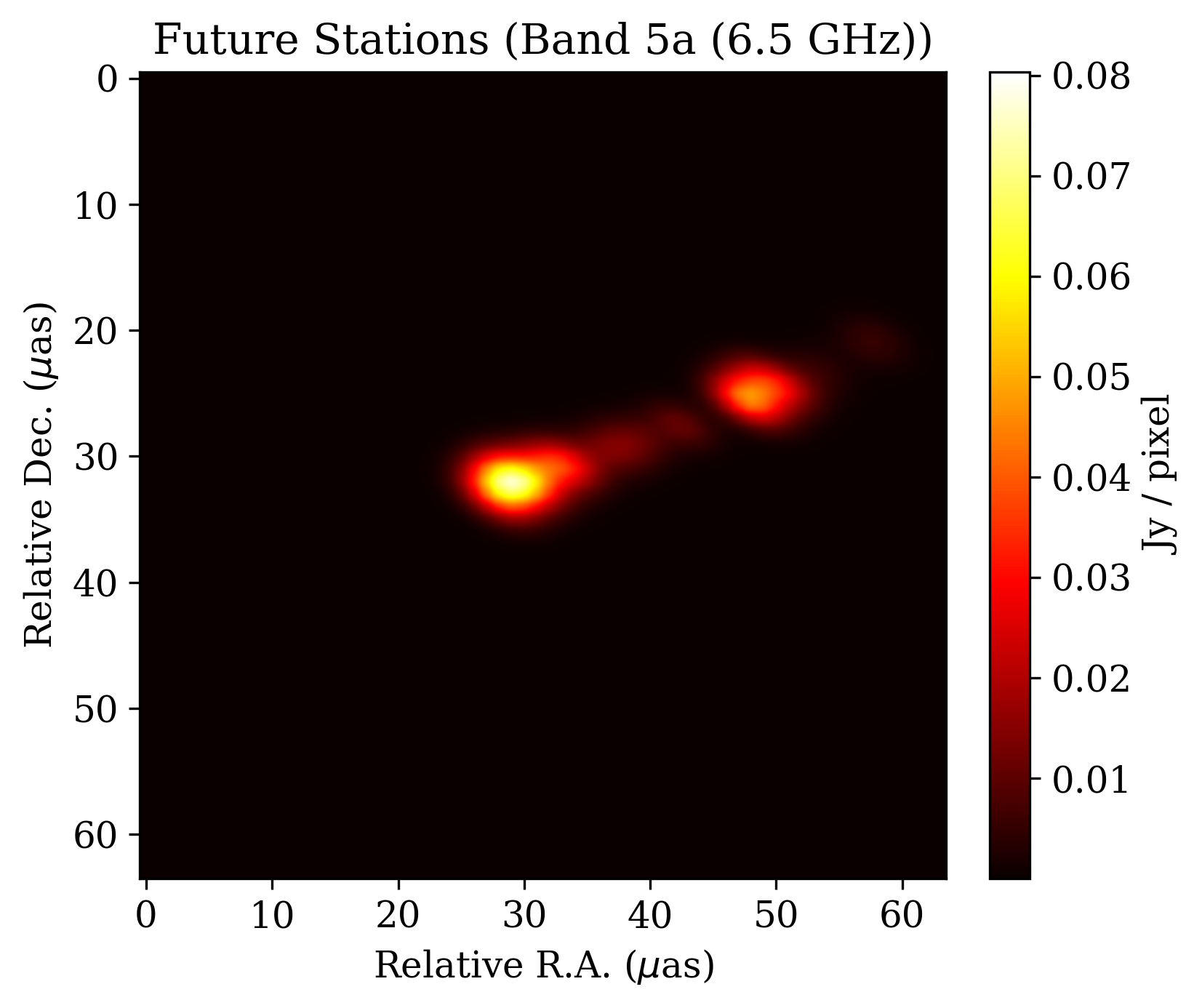}\\
    \caption{First row: \textit{Toy model} of a relativistic jet to simulate a real scientific case with a value for declination of $-30^{\circ}$ and an observing time of $6$ hours. Second row: \emph{uv}-coverage comparison between the Current, Prospective and Future stations scenarios described in the text. Third row: Simulations of the jet with each array operating at SKA band-5a (6.5\,GHz), each labelled correspondingly in the graph's title. The simulations show how the Prospective and Future stations dramatically improve the angular resolution in the observations, with an optimal result in the Future case that allows the diffuse emission of the jet to be detectable.}
\label{fig:uv-coverage2}   
\end{figure*}

\section{Science Communication and Public Engagement in Africa: Bridging Research and Human Capital Development}
\label{sec:SCPE}

The inception of SKA in Africa over a decade ago has not only transformed the radio astronomy landscape of the continent but also changed the scientific narrative of Africa and provided a compelling science communication tool for modern-day outreach, education and training in astronomy and science (e.g. \citealt{atemkeng2020}). Africa has been using public engagement within this framework to serve as a bridge between scientific research, public participation, policymakers, and the development of technical skills. The observatories and techniques being developed in the SKA Partner countries, as well as the spin-offs being felt in other African countries, are tangible anchors for public engagement and offer unique opportunities to bring astronomy to the African audiences.

During the early stages of the SKA and MeerKAT programs, science communication and public engagement became essential to infrastructural development. Events such as National Open Days, documentaries, media engagements, and public exhibitions celebrated technical and technological milestones, including the launch of GRAO, exemplifying the narrative of African science and astronomy capabilities. Training and education programs such as DARA has strategically sought to grow the human capacity development across the SKA Partner countries while simultaneously showcasing their achievements in national media and public events — contributing to public engagement and shaping the narrative of astronomy and astronomy education in Africa and around the world. 

Several flagship events by the IAU through its Office of Astronomy for Development (OAD), Office for Astronomy Outreach (OAO), and Office for Astronomy Education (OAE) have supported astronomy development projects, astrotourism initiatives, public engagement activities, teacher training programs in astronomy, etc. across continental Africa.
Long-term community programs supported by SARAO at the Karoo Desert sites have also continued to combine bursary schemes, local business inclusion, and education roadshows — linking high-tech science to tangible local benefits. These initiatives have greatly impacted astronomy outreach in Africa over the long term.

The African Astronomical Society (AfAS) has also become a professional anchor for public engagement, outreach and human capacity coordination in the African astronomy space, with its vision to
“create and support a globally competitive and collaborative astronomy community in Africa” and “develop astronomy and human capacity throughout the continent” (\citealt{AfAS2024}). The \citet{OAD2025} annual reports recorded that over two million people have been reached globally through its astronomy for development projects, with at least five of the 2025 funded projects based in Africa focusing on public engagement initiatives such as astrotourism, astronomy for mental health, and STEM education and empowerment \citep{Nyangi2025}. The efforts of the OAO NOCs have also been increasingly multilingual and inclusive, with events including livestreams across languages and regions as well as planetarium shows, night sky viewing events, BLV and interactive activities organised for audiences of all ages, educational backgrounds and experience (\citealt{OAO2025}). The synergy between AfAS's coordination role, such as the Cascade Outreach Program, OAD's development model, OAO's NOCs activities, and SKAO's human capital development plans (\citealt{SKAO2024}) shows a growing alignment between research infrastructure and public engagement strategy and the growing impact of these efforts is measurable across several dimensions.

However, there are still gaps remaining; outreach and public engagement activities are unevenly distributed across Africa due to challenges such as language barriers, limited resources and infrastructure, and low or no internet connectivity across the continent, which constrains participation by rural and underserved communities. Many programs and projects remain short-term or under-evaluated due to a lack of or insufficient funding and support from local institutions and the government. The next phase of public engagement in Africa will require a consolidated continent-wide public engagement action plan that links outreach events across not only the SKA African Partner countries but also all of Africa. Such initiatives will need to include multilingual resources and kits curated and co-produced with teachers, astronomers, and institutions, as well as training programs tailored to different learning levels and interests. Additionally, high-level dialogue with policymakers and government institutions is necessary to change the narrative of Africa’s place in global astronomy. Leveraging AfAS' network, IAU Offices' engagement frameworks, and SKAO’s technical expertise will ensure that science communication delivers both research visibility and human capital development on the continent.

Ultimately, Africa’s AVN–SKA journey has shown how science communication, when designed as a developmental system, can translate cutting-edge research into inclusive knowledge communities capable of sustaining global scientific leadership.

\section{Summary, Conclusion \& Recommendations}
\label{sec:SCR}

In this chapter, we have presented the motivation and potential framework for establishing a continental African Very Long Baseline Interferometry (VLBI) network that could operate in synergy with SKA-Mid. Building upon the experience of HartRAO and GRAO, we propose that installing a small number of SKA-Mid–type antennas across Africa would create a scientifically powerful and technically coherent facility. Such an array would fill critical gaps in global VLBI coverage — particularly along the North–South axis — while providing intermediate baselines that directly complement SKA-Mid.

The scientific impact of such a network would be considerable. It would enhance angular resolution and imaging quality for VLBI observations of compact radio sources, expand access to southern sky targets, and strengthen global VLBI collaborations. The use of SKA-Mid receivers, including Band 1, would also extend VLBI capability into a frequency range that links low-frequency SKA-Low science with higher-frequency VLBI, creating a more continuous and versatile observing framework.

\newpage
\begin{landscape}
\begin{table}

\scriptsize
\setlength{\tabcolsep}{1pt}
\renewcommand{\arraystretch}{1}
\rowcolors{3}{black!2}{white}

\begin{adjustbox}{max width=\textheight}
\begin{tabularx}{\textwidth}{L*{27}{C}}
\cmidrule(lr){1-28}
& \multicolumn{27}{c}{\textbf{Stations}}\\
\cmidrule(lr){1-28}
& \multicolumn{1}{c}{\rotH{Algeria}}
& \multicolumn{1}{c}{\rotH{Benin}}
& \multicolumn{1}{c}{\rotH{Botswana$^{\star}$}}
& \multicolumn{1}{c}{\rotH{Cameroon}}
& \multicolumn{1}{c}{\rotH{Congo}}
& \multicolumn{1}{c}{\rotH{Congo\_DR}}
& \multicolumn{1}{c}{\rotH{Egypt}}
& \multicolumn{1}{c}{\rotH{Ethiopia}}
& \multicolumn{1}{c}{\rotH{Gabon}}
& \multicolumn{1}{c}{\rotH{Ghana$^{\star}$}}
& \multicolumn{1}{c}{\rotH{Kenya$^{\star}$}}
& \multicolumn{1}{c}{\rotH{Madagascar$^{\star}$}}
& \multicolumn{1}{c}{\rotH{Malawi}}
& \multicolumn{1}{c}{\rotH{Mauritius$^{\star}$}}
& \multicolumn{1}{c}{\rotH{Morocco}}
& \multicolumn{1}{c}{\rotH{Mozambique$^{\star}$}}
& \multicolumn{1}{c}{\rotH{Namibia$^{\star}$}}
& \multicolumn{1}{c}{\rotH{Niger}}
& \multicolumn{1}{c}{\rotH{Nigeria}}
& \multicolumn{1}{c}{\rotH{SKA}}
& \multicolumn{1}{c}{\rotH{Senegal}}
& \multicolumn{1}{c}{\rotH{South\_Africa}}
& \multicolumn{1}{c}{\rotH{Togo}}
& \multicolumn{1}{c}{\rotH{Tunisia}}
& \multicolumn{1}{c}{\rotH{Uganda}}
& \multicolumn{1}{c}{\rotH{Zambia$^{\star}$}}
& \multicolumn{1}{c}{\rotH{Zimbabwe}}\\
\cmidrule(lr){1-28}
Algeria      & \ldots & 16 & 34 & 18 & 23 & 23 & 13 & 24 & 19 & 16 & 27 & 36 & 32 & 40 & 3 & 36 & 32 & 12 & 15 & 36 & 14 & 35 & 15 & 4 & 25 & 30 & 32 \\
Benin        & 16 & \ldots & 20 & 5 & 9 & 9 & 19 & 19 & 5 & 1 & 19 & 27 & 21 & 31 & 15 & 23 & 17 & 3 & \ldots & 22 & 11 & 22 & \ldots & 17 & 16 & 18 & 20 \\
Botswana$^{\star}$     & 34 & 20 & \ldots & 17 & 12 & 12 & 29 & 19 & 16 & 21 & 14 & 11 & 6 & 16 & 34 & 3 & 4 & 24 & 21 & 4 & 30 & 1 & 21 & 33 & 14 & 5 & 4 \\
Cameroon     & 18 & 5 & 17 & \ldots & 5 & 5 & 17 & 15 & 2 & 6 & 14 & 22 & 16 & 27 & 18 & 19 & 14 & 7 & 4 & 19 & 16 & 18 & 6 & 17 & 11 & 13 & 15 \\
Congo        & 23 & 9 & 12 & 5 & \ldots & \ldots & 20 & 14 & 4 & 10 & 11 & 18 & 12 & 23 & 23 & 15 & 10 & 12 & 9 & 14 & 20 & 13 & 10 & 22 & 10 & 9 & 11 \\
Congo\_DR    & 23 & 9 & 12 & 5 & \ldots & \ldots & 20 & 14 & 4 & 10 & 11 & 18 & 12 & 23 & 23 & 14 & 10 & 12 & 9 & 14 & 20 & 13 & 10 & 22 & 10 & 9 & 11 \\
Egypt        & 13 & 19 & 29 & 17 & 20 & 20 & \ldots & 12 & 19 & 20 & 17 & 28 & 25 & 30 & 16 & 30 & 29 & 16 & 18 & 32 & 24 & 30 & 19 & 9 & 16 & 24 & 25 \\
Ethiopia     & 24 & 19 & 19 & 15 & 14 & 14 & 12 & \ldots & 16 & 21 & 5 & 16 & 13 & 18 & 26 & 19 & 20 & 19 & 19 & 23 & 29 & 19 & 20 & 20 & 5 & 14 & 15 \\
Gabon        & 19 & 5 & 16 & 2 & 4 & 4 & 19 & 16 & \ldots & 6 & 14 & 22 & 16 & 27 & 20 & 18 & 13 & 8 & 5 & 18 & 16 & 17 & 6 & 19 & 12 & 13 & 15 \\
Ghana$^{\star}$        & 16 & 1 & 21 & 6 & 10 & 10 & 20 & 21 & 6 & \ldots & 20 & 28 & 22 & 33 & 15 & 24 & 18 & 4 & 2 & 22 & 10 & 22 & \ldots & 17 & 18 & 19 & 21 \\
Kenya$^{\star}$        & 27 & 19 & 14 & 14 & 11 & 11 & 17 & 5 & 14 & 20 & \ldots & 11 & 8 & 15 & 28 & 13 & 15 & 20 & 18 & 18 & 29 & 14 & 19 & 24 & 2 & 9 & 9 \\
Madagascar$^{\star}$   & 36 & 27 & 11 & 22 & 18 & 18 & 28 & 16 & 22 & 28 & 11 & \ldots & 6 & 5 & 38 & 8 & 15 & 29 & 27 & 14 & 37 & 10 & 28 & 34 & 13 & 10 & 8 \\
Malawi       & 32 & 21 & 6 & 16 & 12 & 12 & 25 & 13 & 16 & 22 & 8 & 6 & \ldots & 12 & 33 & 5 & 10 & 23 & 21 & 10 & 31 & 6 & 22 & 30 & 9 & 3 & 2 \\
Mauritius$^{\star}$    & 40 & 31 & 16 & 27 & 23 & 23 & 30 & 18 & 27 & 33 & 15 & 5 & 12 & \ldots & 41 & 13 & 20 & 33 & 31 & 18 & 41 & 15 & 32 & 37 & 17 & 15 & 13 \\
Morocco      & 3 & 15 & 34 & 18 & 23 & 23 & 16 & 26 & 20 & 15 & 28 & 38 & 33 & 41 & \ldots & 36 & 32 & 12 & 15 & 36 & 11 & 35 & 15 & 7 & 26 & 31 & 33 \\
Mozambique$^{\star}$   & 36 & 23 & 3 & 19 & 15 & 14 & 30 & 19 & 18 & 24 & 13 & 8 & 5 & 13 & 36 & \ldots & 8 & 26 & 23 & 6 & 33 & 2 & 24 & 34 & 14 & 6 & 4 \\
Namibia$^{\star}$      & 32 & 17 & 4 & 14 & 10 & 10 & 29 & 20 & 13 & 18 & 15 & 15 & 10 & 20 & 32 & 8 & \ldots & 21 & 18 & 4 & 26 & 5 & 18 & 31 & 15 & 6 & 7 \\
Niger        & 12 & 3 & 24 & 7 & 12 & 12 & 16 & 19 & 8 & 4 & 20 & 29 & 23 & 33 & 12 & 26 & 21 & \ldots & 3 & 25 & 10 & 25 & 3 & 13 & 18 & 20 & 22 \\
Nigeria      & 15 & \ldots & 21 & 4 & 9 & 9 & 18 & 19 & 5 & 2 & 18 & 27 & 21 & 31 & 15 & 23 & 18 & 3 & \ldots & 22 & 11 & 22 & 1 & 16 & 16 & 18 & 20 \\
SKA          & 36 & 22 & 4 & 19 & 14 & 14 & 32 & 23 & 18 & 22 & 18 & 14 & 10 & 18 & 36 & 6 & 4 & 25 & 22 & \ldots & 31 & 4 & 23 & 35 & 18 & 9 & 8 \\
Senegal      & 14 & 11 & 30 & 16 & 20 & 20 & 24 & 29 & 16 & 10 & 29 & 37 & 31 & 41 & 11 & 33 & 26 & 10 & 11 & 31 & \ldots & 31 & 10 & 18 & 27 & 28 & 30 \\
South\_Africa & 35 & 22 & 1 & 18 & 13 & 13 & 30 & 19 & 17 & 22 & 14 & 10 & 6 & 15 & 35 & 2 & 5 & 25 & 22 & 4 & 31 & \ldots & 22 & 34 & 14 & 5 & 4 \\
Togo         & 15 & \ldots & 21 & 6 & 10 & 10 & 19 & 20 & 6 & \ldots & 19 & 28 & 22 & 32 & 15 & 24 & 18 & 3 & 1 & 23 & 10 & 22 & \ldots & 17 & 17 & 19 & 21 \\
Tunisia      & 4 & 17 & 33 & 17 & 22 & 22 & 9 & 20 & 19 & 17 & 24 & 34 & 30 & 37 & 7 & 34 & 31 & 13 & 16 & 35 & 18 & 34 & 17 & \ldots & 22 & 29 & 30 \\
Uganda       & 25 & 16 & 14 & 11 & 10 & 10 & 16 & 5 & 12 & 18 & 2 & 13 & 9 & 17 & 26 & 14 & 15 & 18 & 16 & 18 & 27 & 14 & 17 & 22 & \ldots & 9 & 9 \\
Zambia$^{\star}$       & 30 & 18 & 5 & 13 & 9 & 9 & 24 & 14 & 13 & 19 & 9 & 10 & 3 & 15 & 31 & 6 & 6 & 20 & 18 & 9 & 28 & 5 & 19 & 29 & 9 & \ldots & 2 \\
Zimbabwe     & 32 & 20 & 4 & 15 & 11 & 11 & 25 & 15 & 15 & 21 & 9 & 8 & 2 & 13 & 33 & 4 & 7 & 22 & 20 & 8 & 30 & 4 & 21 & 30 & 9 & 2 & \ldots \\
\cmidrule(lr){1-28}
\end{tabularx}
\end{adjustbox}

\vspace{-0.1in}

\caption{The proposed continental African VLBI network stations by country — baseline lengths are in units of $10^{7}\lambda$ (where; $\lambda$ $\approx$ 5\,cm). \\$^{\star}$\,denote original SKA African Partner country.  Our motivation for the choice of these additional dozen and a half countries is to allow\\ for good geographical distribution of the array coupled with the fact that each one of these countries currently host, or have hosted legacy\\ satellite Earth-station antennas in the past.}.
\label{tab:avn_baseline}

\end{table}
\end{landscape}

\newpage

At present, the original SKA African Partner countries concept — based on refurbishing legacy telecommunication antennas — has proven extremely challenging to realise. The technical complexity, high refurbishment costs, and limited local expertise in many partner countries have slowed progress. A shift toward constructing new SKA-standard antennas offers a more sustainable and scientifically strategic path forward. Nevertheless, this approach will still require careful planning, substantial investment in data infrastructure, and ongoing development of human and technical capacity across the continent.

We therefore recommend that future efforts focus on:

\begin{itemize}
\item Assessing the feasibility of deploying new SKA-Mid–compatible antennas in selected African partner countries to form an intermediate-baseline VLBI array.
\vspace{-0.075in}
\item Developing a coordinated technical and funding strategy for data transport, correlation, and long-term operational sustainability within the broader SKA framework.
\vspace{-0.075in}
\item Strengthening capacity-building programmes in radio astronomy, engineering, and data science to ensure local participation in both operations and scientific exploitation.
\vspace{-0.075in}
\item Encouraging partnerships between African institutions, SKA Observatory, and international VLBI networks to align technical standards and scientific goals.
\end{itemize}

While the concept outlined here remains a proposal, it represents a compelling vision for the next phase of Africa’s engagement in global radio astronomy. By building upon existing momentum around SKA-Mid and leveraging the continent’s geographic advantage, a coordinated African VLBI network could both advance frontier science and catalyse technological and educational development across the region.
\vspace{0.2in}

{\bf Acknowledgement}:
 We acknowledge financial support by the German Federal Ministry of Education and Research (BMBF) under ErUM-Pro grant 05A23WW3 (Verbundprojekt D-MeerKAT III). We thank the anonymous referee for their detailed comments which helped to improve this paper.
\vspace{0.2in}

\bibliographystyle{abbrvnat-maxbibnames4}
\bibliography{chapter} 

\end{document}